\documentclass[envcountsame]{llncs}
 \pdfoutput=1

\usepackage{xspace}
\usepackage{latexsym}
\usepackage{amssymb}   %

\usepackage{ifthen}
\newboolean{commentsaon}         %
\setboolean{commentsaon}{true}
  \setboolean{commentsaon}{false}  %

\newboolean{commentson} %
\setboolean{commentson}{true}

\newcommand{\comment}[1]
{\ifthenelse{\boolean{commentson}\AND\boolean{commentsaon}}
   {{\par\noindent\mbox{}{\small\blue[ *** #1 ]\par}\noindent\par}}{}}

\newcommand{\commenta}[1]
{\ifthenelse{\boolean{commentsaon}}
   {{\par\noindent\mbox{}{\small\color[rgb]{0, .5, 0}[ *** #1 ]\par}\noindent\par}}{}}

\usepackage{pgf}

\usepackage[yyyymmdd]{datetime}

\newcommand{\myhalfmagnification}{0.01}

\usepackage{color}
\newcommand\blue     {\color{blue}}

\usepackage{hyperref}
\usepackage{pgf}

\newcommand*{\seq}[2][n]  {{#2_{1}, \allowbreak \ldots, \allowbreak #2_{#1}}}

\newcommand*{\TU}{{\ensuremath{\cal{T U}}}\xspace}
\newcommand*{\TB}{{\ensuremath{\cal{T B}}}\xspace}

\usepackage[mathscr]{euscript} 

\newcommand*{\OO}{{\ensuremath{\cal O}}\xspace}

\newcommand{\Var}{{{\ensuremath{\mathcal{V}\!{\it a r}}}\xspace}}

\newcommand*{\NN}{{\ensuremath{\mathbb{N}}}\xspace}

\newcommand*{\Tpi}[1][]{{\ensuremath{\mbox{\bf T}^\pi_{#1}}}}
\newcommand*{\TpiP}{{\ensuremath{\mbox{\bf T}^\pi_P}}\xspace}

\newcommand*{\myunderscore}{\mbox{\tt\symbol{95}}}

\newcommand*{\nqueens}{\mbox{\sc nqueens}\xspace}

\begin{document}

\title{S-semantics -- an example}

\author   {W{\l}odzimierz Drabent%
           \\[1ex]
             {\small \today%, Version 1.8%
             }%
} %

 \institute{
       \normalsize\small
      \begin{tabular}{c}
           Institute of Computer Science,
           Polish Academy of Sciences,\\
          and \\
           Department of Computer and Information Science,
           Link\"oping University
           \\[.5ex]
           \texttt{drabent\,{\it at}\/\,ipipan\,{\it dot}\/\,waw\,{\it dot}\/\,pl}
      \end{tabular}
 }

 \maketitle 
\noindent

\begin{abstract}
     The s-semantics makes it possible to explicitly deal with
     variables in program answers. So it seems suitable for programs
     using nonground data structures, like open lists. However it is
     difficult to find published examples of using the s-semantics to reason
     about particular programs.

     Here we apply s-semantics to prove correctness and completeness
of Fr\"uhwirth's $n$
     queens program. This is compared with a proof, published elsewhere,
     based on the standard semantics and Herbrand interpretations.
    
\end{abstract}

\section{Introduction}
The s-semantics for definite logic programs
\cite{DBLP:journals/tcs/FalaschiLPM89,s-semantics94,DBLP:journals/tcs/Bossi09}
deals explicitly with variables in program answers.
So such semantics may seem suitable for reasoning about programs which use
nonground data structures, like open lists.   
This paper applies the s-semantics to establish correctness and completeness
of the $n$ queen program of Fr\"uhwirth  \cite{fruehwirth91}.
The program uses open lists with possibly nonground members.
Due to the importance of
nonground data structures for the program, it may even seem that the
standard semantics is not sufficient here.
  This is not the case, another paper \cite{drabent.nqueens.tplp.pre}
presents correctness and
completeness proofs for the program, based on Herbrand interpretations and
the standard semantics.
So those proofs can be compared with the ones presented here.
Maybe surprisingly,
it turns out that the standard semantics is preferable, as
it leads to substantially simpler specifications and proofs.
It should be added that many ideas from \cite{drabent.nqueens.tplp.pre} are
used in this paper.

  It is difficult to find applications of s-semantics to reasoning
  about particular programs.  (The author is not aware of any.)
  Thus the proofs presented here
  provide a, hopefully useful, example.

The paper is organized as follows.  This introduction is concluded with
preliminaries. 
The next two sections present, respectively, the s-semantics 
(together with sufficient conditions for correctness and completeness)
and the $n$
queens program.  Section \ref{sec.correctness} discusses correctness of the
program, first
constructing a specification for correctness, then presenting a correctness
proof.  Section \ref{sec.completeness} discusses completeness in a similar
way.  The last section summarizes the paper.

\paragraph{Preliminaries}
This paper considers definite clause logic programs.
It uses the standard notation and terminology, 
following
\cite{Apt-Prolog}.  So we deal with queries (conjunctions of atoms) instead
of goals.  We assume a fixed alphabet (of predicate and function symbols, and
variables).
The set of variables will be denoted by \Var,
the set of terms (over the alphabet) by \TU, and the set of atoms by \TB;
\NN stands for the set of natural numbers.
Given a program $P$,
a query $Q$ such that $P\models Q$ is called an {\em answer} (or correct
answer) of $P$.
We will use answers,
to avoid dealing with computed (or correct) answer substitutions.
(In \cite{Apt-Prolog}, answers are called correct instances of queries.)
By a {\em computed} (or SLD-computed) {\em answer} $Q'$ for a query $Q$ we mean an
answer obtained by 
means of SLD-resolution (so $Q'$ is a computed instance \cite{Apt-Prolog}
of $Q$, in other words
$Q'=Q\theta$ for a computed answer substitution $\theta$).
By the {\em relation} defined by a predicate $p$ in $P$ we mean
    $\{\,\vec t\,\in\TU^n \mid P\models p(\vec t\,) \,\}.\!$%

An expression (term, atom, sequence of terms, etc) is {\em linear} if no
variable occurs in it twice.
Expressions $\seq E$ ($n>0$) are {\em variable disjoint} if for each
$0<i<j\leq n$ no variable occurs in both $E_i$ and $E_j$.
As in Prolog, each occurrence of \myunderscore\ in an expression denotes a
distinct variable.

We use the standard list notation of Prolog.
An {\em open list} (a {\em list}) of length $n\geq0$
is a term $[\seq t|v]\in\TU$ where $v\in\Var$
(resp.\ $[\seq t]\in\TU$);
$v$ is the {\em open list variable} of $[\seq t|v]$.
The term $t_i$ ($0<i\leq n$) is called the $i$-th {\em member} of the (open)
list. 
For $n=0$, $[\seq t|t]$ stands for $t$.
So an empty open list (i.e.\ of length 0) is a variable.
The tail of a list $l$ will be denoted by
t{}l$(l)$, so ${\rm t l}([t|u])=u$.  By the tail of
an empty open list, ${\rm t l}(\myunderscore)$ we mean a new variable,
distinct from any other variable in the context.

\section{S-semantics}

The s-semantics \cite{DBLP:journals/tcs/FalaschiLPM89}
 was introduced to capture the phenomenon that logically
equivalent programs may have distinct sets of computed answers for a given
query.  Consider an example \cite{DM87short,DM88} of two programs%
\[
\begin{array}[t]{l}
  p(f(X)). \\
  p(f(a)). \\
\end{array}
\hspace{7em}
\begin{array}[t]{l}
  p(f(X)). \\
\end{array}
\]
They are logically equivalent, have the same set of logical consequences (thus
the same set of answers), and have the same least Herbrand model
(for any alphabet containing $p,{f},a$).
However for a query $p(Y)$, the SLD-resolution
produces two answers for the first program, while only one answer is produced
for the second one.  (The answer $p(f(a))$ is not produced.)%
\footnote{%
  The observation (that logically equivalent programs may have distinct sets of
  computed answers for the same query) is in 
  \cite[Section 3.1]{DBLP:journals/tcs/Bossi09} incorrectly attributed to
  \cite{DBLP:journals/tcs/FalaschiLPM89}. 
  However, it was presented in Pisa in 1987 \cite{DM87short}.  The
  author is not aware of any earlier appearance of such observation.
} %

The s-semantics captures such differences by describing the answers
produced for most general atomic queries.
\begin{definition}[S-semantics]\rm
\nopagebreak
  Let $P$ be a program.  Its s-semantics is given by the set
  \[
  \OO(P) = \left\{\,  A\in \TB\: \left|
      \begin{tabular}{l}
        $A$ is an SLD-computed answer \\ for a query $p(\seq V)$, where \\
        $p$ is a predicate symbol of arity $n$, \\
         and
        $\seq V$ are distinct variables
      \end{tabular}
      \right\}\right..
  \]
In other words, $A=p(\seq V)\theta$ where $\theta$ is an SLD-computed answer
substitution for query  $p(\seq V)$.
\end{definition}

We use here a slight simplification of the original s-semantics.  There,
the members of $\OO(P)$ are not atoms but equivalence classes of atoms under
the equivalence relation $\approx$ of variable renaming.%
\footnote{%
Both version are equivalent.  Let $\OO'(P)$ be the original s-semantics of
$P$.  Then $\OO(P) = \bigcup{\OO'(P)}$, and $\OO'(P)$ is the quotient set
$\OO(P)/_\approx$ of $\OO(P)$ w.r.t.\ $\approx$.
}
Obviously, the set of ground instances of $\OO(P)$ is the least Herbrand
model of $P$.
This is a main property of the s-semantics:
\begin{lemma}\rm
\label{lemma.s-semantics}
  Let $P$ be a program.
A query $Q=\seq B$ has an SLD-computed answer $Q'$ iff 
there exist $\seq{A}\in\OO(P)$ such that

\quad
the $n+1$ expressions $Q,\seq {A}$ are variable disjoint, 

\quad
$Q'=Q\gamma$ for an mgu $\gamma$ of $Q$ and $\seq {A}$.
\end{lemma}

The s-semantics is the $\subseteq$-least fixed point of a 
specific
immediate consequence operator.

\begin{definition}\rm
The s-semantics {\em immediate consequence operator} for a program $P$ is the
function $\TpiP\colon 2^\TB\to 2^\TB$ defined by
{\sloppy\par}
\[
 \TpiP(I) = \left\{\,  H\theta\: \left|
    \mbox{%
      \begin{tabular}{l@{}}
        $\theta$ is an mgu of $(\seq B)$ and  $(\seq A)$ \\
        for some
        $(H\mathop\gets\seq B) \in P$, \ 
        $\seq A\in I$ such  \\
        that $\seq A$,\,$(H\gets\seq B)$ are variable disjoint \\
      \end{tabular}%
      } %
      \right\}\right..
\]
\end{definition}
In the definition and in Lemma \ref{lemma.s-semantics}
 it is important that $n+1$ expressions are pairwise variable disjoint.%
\footnote{%
The wording used in
\cite{DBLP:journals/tcs/FalaschiLPM89,s-semantics94,DBLP:journals/tcs/Bossi09}
may be incorrectly understood as requiring that
$(H\gets\seq B)$ is variable disjoint with 
$(\seq A)$.
Cf.\ e.g.\,``[atoms] are renamed apart w.r.t.\ the clause''
in the definition  of  \TpiP in
 \cite[p.\,4696]{DBLP:journals/tcs/Bossi09}.
}
Also, note that an mgu of two ground expressions is any renaming substitution.

For any $I\subseteq\TB$, $\TpiP(I)$ is closed under variable renaming
(as for any renaming $\gamma$, if $\theta$ is an mgu of $\vec B$ and $\vec A$
then $\theta\gamma$ is an mgu of $\vec B$ and $\vec A$ too
\cite[Lemma\,2.23]{Apt-Prolog}).
The operator is continuous in the lattice $(2^\TB,\subseteq)$,
its least fixed point is $(\TpiP)^\omega(\emptyset)$,  and we have
\[
\OO(P) = (\TpiP)^\omega(\emptyset).
\]

By a {\bf specification} (for s-semantics) we mean a set $S\subseteq\TB$, a program
$P$ is {\bf correct} w.r.t.\ $S$ when $\OO(P)\subseteq S$.
Here are sufficient conditions for correctness.
\begin{theorem}[Correctness]\rm
  \label{theorem.correctness}
  Let $P$ be a program and $S\subseteq\TB$.

  If $\TpiP(S)\subseteq S$ then $\OO(P)\subseteq S$.

  If $\Tpi[\!\{C\}](S)\subseteq S$ for each clause $C\in P$ then $\OO(P)\subseteq S$.
\end{theorem}
{\sc Proof }
The least fixed point $\OO(P)$ of $\TpiP$ is the least $I\subseteq\TB$
such that $\TpiP(I)\subseteq I$.
$\TpiP(I)= \bigcup_{C\in P}\Tpi[\!\{C\}](I)$, thus the premises of both
implications are equivalent. $\Box$

\smallskip\smallskip
The notion of correctness in logic programming differs from that in 
imperative and functional programming.  Due to the nondeterministic
nature of logic programming,
it is not sufficient that a program is correct; e.g.\ the empty program is
correct w.r.t.\ any specification.
We also need that the program produces the
required answers; we are interested in program completeness.
A program $p$ is {\bf complete} w.r.t.\ a specification $S$ when
$S\subseteq\OO(P)$.

To deal with completeness, let us introduce an auxiliary notion.
By a {\bf level mapping}
we mean a function $|\ |\colon S\to \NN$
assigning natural numbers to atoms from a set $S\in\TB$,
such that if $A,A'\in\TB$ are variants then $|A|=|A'|$.
(Note that usually one considers level mappings defined on ground atoms
\cite{Apt-Prolog}.)

\pagebreak[3]
\begin{theorem}[Completeness]\rm
\label{theorem.completeness}
Let $P$ be a finite program and $S\subseteq\TB$.  
Assume that there exists a level mapping $|\ |\colon S\to\NN$ such that for each
$A\in S$
\\
\mbox{\quad\ }%
$A\in \Tpi[C](\{\seq{A}\})$ for some clause $C\in P$
and some  $\{\seq{A}\}\subseteq S$
\\
\mbox{\quad\ }%
where $|A| > |A_i|$ for $i=1,\ldots,n$.
 \\
 Then $S\subseteq \OO(P)$.
\end{theorem}
It is sufficient to consider only such $\seq A$
that are variable disjoint and $n$ is the number of body atoms in $C$.
As $S$ may be not closed under renaming, it is sometimes useful to generalize 
condition ``$\{\seq{A}\}\subseteq S$'' to ``$\seq A$ are variants of some
atoms from $S$\,''.

\medskip
\noindent
{\sc Proof} (of the more general version of Theorem \ref{theorem.completeness}) \
By induction on $i$ we show that
 $S_i=\{ A\in S \mid |A| < i \,\}\subseteq (\TpiP)^i(\emptyset)$. 

For $i=0$ the thesis holds vacuously.
Assume that it holds for some $i\in\NN$ and consider an $A\in S_{i+1}$.
For some clause $C\in P$ we have $A\in \Tpi[C](\{\seq{A}\})$, where
for $k=1,\ldots,n$ atom $A_k$ is a variant of some $A_k'\in S$
and $i\mathop+1 \mathop>|A|  \mathop> |A_k| \linebreak[3]
\mathop=|A_k'|$. %
Hence $A_k'\in S_i$ and, by the inductive assumption, 
$A'_k,A_k\in (\TpiP)^i(\emptyset)$.  As  $A\in \Tpi[C](\{\seq{A}\})$, we have
$A\in (\TpiP)^{j+1}(\emptyset)$.
$\Box$

\medskip
The sufficient conditions for correctness and completeness
of Theorems \ref{theorem.correctness}, \ref{theorem.completeness}
are similar to those related to the standard semantics
\cite{Clark79,Deransart.Maluszynski93}, 
(see \cite{Drabent.tocl16} for comments and references).

{\sloppy\par}

\section{The $n$ queens program}
\label{sec.program}

Thom Fr\"uhwirth presented a short, elegant and efficient Prolog program for
 the n-queens problem \cite{fruehwirth91}.
 However the program may be seen as rather tricky and one may be not convinced
 about its correctness. 
We apply the s-semantics to prove its correctness and completeness.
This section, based on \cite{drabent.nqueens.tplp.pre}, presents the program
and introduces some notions used later in the specifications and proofs.

The problem is to place $n$ queens on an $n\times n$ chessboard, so that no
two queens are placed on the same row, column, or diagonal.
The main idea of the program is to describe the placement of the queens by a
data structure in which it is impossible that two queens 
violate the restriction
(there are some exceptions, this will be clear later on).
In this way all the constraints of the
problem are treated implicitly and efficiently.
Here is the program, in its simplest version not using Prolog arithmetic,
with predicate names abbreviated 
($q u$ for
{\small${\tt queens p}$},
$g l$ for {\small${\tt gen\myunderscore list p}$},
$p q$ for {\small{\tt place\myunderscore queen}},
and ${\it p q s}$ for %
{\small${\tt place\myunderscore queens p}$}).
\begin{eqnarray}
&& \nonumber
      q u(N,Q s) \gets g l(N,Q s), p q s(N,Q s,\myunderscore,\myunderscore).
      \\
&& \nonumber   g l (0,[\,]).   \\
&& \nonumber   g l(s(N),[\myunderscore|L]) \gets
            g l(N,L).
\\
\label{clause1}
&&    p q s(0,\myunderscore,\myunderscore,\myunderscore).        \\
\label{clause2}
&& \it   p q s(s(I),Cs,Us,[\myunderscore|D s]) \gets
    \begin{array}[t]{l}
   \it         p q s(I,Cs,[\myunderscore|Us],D s), \\
   \it         p q(s(I),Cs,Us,D s).  
    \end{array}
\pagebreak[3]
\ifthenelse{\boolean{commentson}\AND\boolean{commentsaon}}{}{\pagebreak}
\\
 && \nonumber
 \mbox{\small\%\ 
       $\it p q(Queen,Column,Up diagonal,Down diagonal)$ places a single queen}
  \\
\label{clause3}
&& \it   p q(I,[I|\myunderscore],[I|\myunderscore],[I|\myunderscore]).        \\
\label{clause4}
&& \it p q(I,[\myunderscore|Cs],[\myunderscore|Us],[\myunderscore|D s]) \gets
            p q(I,Cs,Us,D s).
\end{eqnarray}
Its main predicate $q u$ provides solutions to the problem,
in an answer $q u(n,q s)$, $n$ is a number and $q s$ encodes a solution as 
a list of length $n$.
The interesting part of the program consists of clauses
(\ref{clause1}),\ldots,(\ref{clause4}).
So this fragment is our program of interest, it will be called \nqueens.

Solutions to the $n$ queens problem are provided by
the answers of program \nqueens 
of the form $p q s (n, q s, t_1, t_2)$,  where $n>0$ and $q s$ is
a list of length $n$. 
(The remaining arguments may be understood as internal data.)
So an initial query  $p q s (n, q s_0, \myunderscore,\myunderscore)$,
where $q s_0$ is a list of $n$ distinct variables can be used to obtain the
solutions.

To understand a logic program from a declarative point of view we need to
understand the relations defined by the predicates of the program.
This can be done abstracting from any operational semantics.  Such
possibility is an advantage of declarative programming, and of logic
programming in particular.
We first explain the relations informally and then construct a formal
specification.  We begin with discussing the data of the program.

The natural numbers are represented by terms in a standard
way, a number $n$ as $s^n(0)$.
Assume that columns and rows of the chessboard are numbered from the left/top.
Each queen is identified by its row number.
The chessboard is represented as a (possibly) open list, with number $i$
appearing as the $j$-th member when the queen (of row) $i$ is in column $j$.
Empty column $j$ is represented as a variable being the $j$-th member
(or the length of the list being\,${}<j$).

\begin{figure}
\[
  \includegraphics
        [trim = 4.4cm 18.2cm 3.3cm 5.4cm, clip, scale=.9] 
  {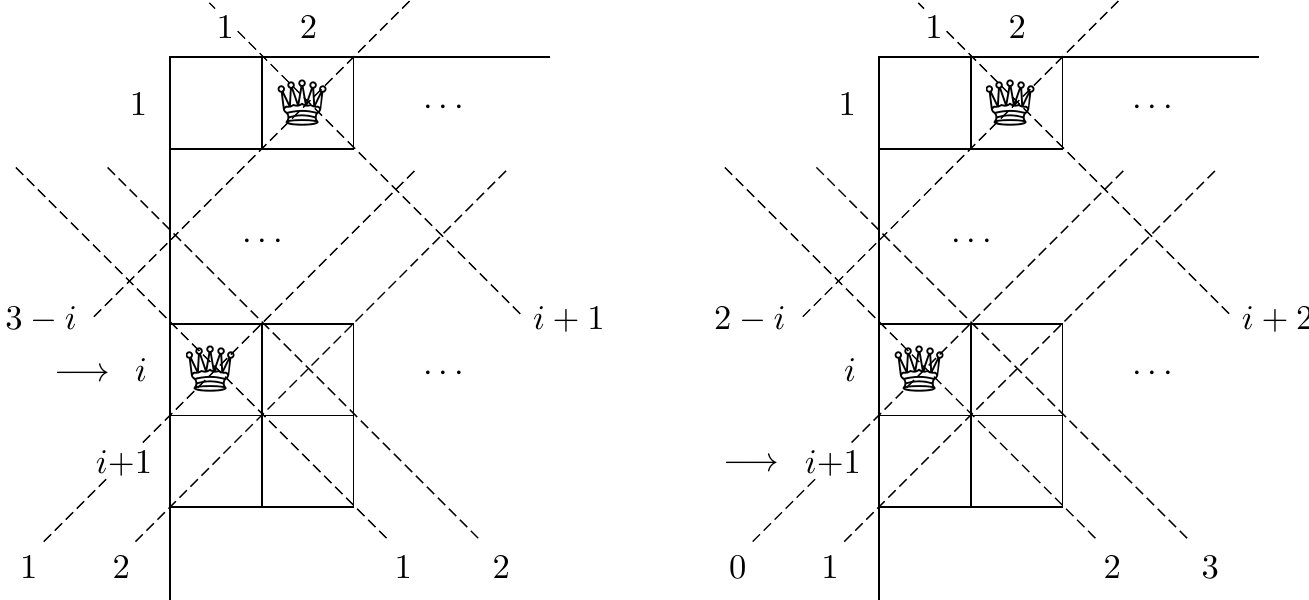}%
\vspace{-3ex plus 3ex}
\]
  \caption[just a trick]{%
 \protect\cite{drabent.nqueens.tplp.pre}
Numbering of rows and columns.  %
Numbering of up  (\rotatebox[origin=c]{-45}{$|$}) diagonals and
down (\rotatebox{45}{$|$}) diagonals
 in the context of row   $i$ (left), and $i+1$ (right).

The board with two queens is represented in the context of row $i$ as follows:
the columns by  $[i,1|\ldots]$, the up diagonals by  $[i|\ldots]$,
the down diagonals by $[i,\ldots,1|\ldots]$ (where $1$ is the member number
$i+1$).
Diagonals with non-positive numbers are not represented.
In the context of row $i+1$, the down diagonals are represented by
 $[t,i,\ldots,1|\ldots]$ (where $1$ is the member number $i+2$, and $t$ is
arbitrary). 
}
\label{figure.diagonals}
\vspace{-1.1ex}
\end{figure}

An up (respectively down) diagonal consists of the fields with the same 
sum (difference) of the row and column number.
Diagonals intersecting a given row are numbered from the left
(Figure \ref{figure.diagonals}).
In contrast to the numbering of rows and columns, 
this numbering is not fixed.
It depends on the context, namely
on which row we focus.
Diagonal $j$ includes the $j$-th field of the row.
Thus, in the context of row number $i$, its queen $i$ is
in the column and in the up and down diagonals of the same number.
The up (the same for down) diagonals are represented by an open list of
numbers, a number $i$ 
as the $j$-th member of the list means that the $j$-th diagonal contains the
queen $i$.  If no queen is placed in the diagonal number $j$, the 
$j$-th member of the list is a variable (or does not exist).
This representation guarantees that at most one queen can be placed in each
column and diagonal (except for those with negative numbers).

Now let us outline (rather superficially) the semantics of \nqueens.
The idea is that $p q$ defines a relation consisting of tuples  
$
(i, cs, us, d s )
$
where $i$ is the number of a row, and $cs, us, d s$ are (possibly open) lists
representing columns, up diagonals, and down diagonals respectively,
and, for some $j>0$, the j-th member of each list is $i$.
(Actually, these tuples are instances of those described by the s-semantics of
\nqueens.)

The relation defined by ${\it p q s}$ consists of tuples
$(0, cs, us, d s )$ (with arbitrary ${\it cs, us, d s}$) and 
$(i, cs, us, [t| d s'] )$, where $i>0$ and
$cs$ describes a placement of queens number $1,\ldots,i$ in the columns, and
$us, d s'$ describe their placement in the diagonals (numbered in the context
of row $i$).   Moreover, in the chessboard fragment of rows $1,\ldots,i$,
each row, each column, and each diagonal contains at most one queen.%
\footnote{%
Notice that the last statement follows from the previous one, but only for
the columns and diagonals represented by $cs, us, d s'$ (i.e.\ 
those intersecting the row $i$).  However there are down diagonals of
numbers $-i+2,\ldots,-1,0$ that intersect some of the rows $1,\ldots,i-1$,
but not row $i$.  
}

\section{Correctness of \nqueens}
\label{sec.correctness}
\subsection{Specification for correctness}

For discussing program correctness it is reasonable to use a specification
which is a suitable superset of the actual semantics $\OO(\nqueens)$ of the
program. 
The specification should imply the program properties of interest.
(More precisely, correctness w.r.t.\ the specification should imply them.)
Also, it is useful when a specification neglects unnecessary
details of the semantics of the program.  This may make simpler both the
specification and the correctness proof.

 Our specification for $p q$ is
\[
S_{p q} = 
\{\,
\begin{array}[t]{@{}l@{}}
  p q(\,v,\, [\seq[k]c,v|c_0],\, [\seq[k]u,v|u_0],\, [\seq[k]d,v|d_0] \,) 
  \in\TB
  \mid{}
  \\
  k\geq0, \  
  v,c_0,\ldots,c_k,u_0,\ldots,u_k,d_0,\ldots,d_k \mbox{ are distinct variables}
  \,\}
\end{array}
\]
Here all the variables occurring in the three open lists are distinct, except
for $v$, which occurs in the three lists at position $k+1$, to represent the
same queen in the column, up diagonal, and down diagonal number $k+1$.

For a formal specification of $p q s$, let us introduce some auxiliary notions.
Assume that a queen $j\in\NN\setminus\{0\}$ 
(i.e.\ the queen of row $j$) is placed in column $k$
(i.e.\ $j$ is the $k$-th member of a possibly open list $cs$ representing
columns). 
Then, in the context of row $i$ (say $i\geq j$), the queen $j$ is on the up
diagonal with number $k+j-i$;
we say $k+j-i$ is the {\bf up diagonal number} of queen $j$ in $cs$ w.r.t.\ $i$
\cite{drabent.nqueens.tplp.pre}.
Similarly, $k+i-j$ is the {\bf down diagonal number} of queen $j$ 
(in $cs$ w.r.t.\ $i$),
as this is the number of its down diagonal in the context of row $i$.
Consider, for instance, the queen $i-3$ placed in column 2.  Then its
up (down) diagonal number w.r.t.\ $i$ is, respectively, $-1$ and $5$.

Writing that some queens have distinct up (or down) diagonal numbers, we will
usually skip ``w.r.t.\,$i$\/'', as the numbers are distinct w.r.t.\ any $i\in\NN$.

We say that a term $t\in\TU$ is a {\bf g.v.d.}\ (ground-or-variable open list
with distinct members) if $t$ 
\vspace{-1ex}
\[
    \begin{tabular}{l}
      is linear,    \\
      is an open list with distinct members, \\
      and each its member is ground or is a variable.
    \end{tabular}
\]
Note that unification of two (unifiable) variable disjoint g.v.d.'s 
which do not have a common ground member results in a g.v.d..

We say that an open list $cs$ {\em represents a correct placement} 
up to row $m$
(in short: is {\bf correct} up to $m$) when $0\leq m$ and
\[
\begin{tabular}{l}
          $cs$ is a g.v.d., \\
          the ground members of $cs$ are $1,\ldots,m$, \\
          their up diagonal numbers in $cs$ are distinct, \\
          their down diagonal numbers in $cs$  are distinct, \\
\end{tabular}
\]

We have to take care that the placement of the queens on the diagonals is
properly reflected in the open lists $us,d s$ representing the diagonals.
Actually, we do not need to specify that $cs,d s$ are open lists.
Let us generalize the notion of list membership:
    A term $s$ is the $k${\bf-th member} of a term $t$ if $t$ is of the form
    $t=[\seq[{k-1}] t,s|t_0]$ (where $0<k$).
    We say that a pair of terms $(us,d s)$ is {\bf correct}
 (represents a correct placement)
up to $m$
w.r.t.\ a row $i\in\NN$ and an (open) list $cs$
when
\[
\begin{tabular}{l}
  for  each  $j\in\{1,\ldots,m\}$, \\
  \quad
  \begin{tabular}{l}
      $j$ is a member of $cs$, and \\
      if
      the up (down) diagonal number of $j$ in $cs$ w.r.t.\ $i$
       is $l>0$
          \\
      then the $l$-th member of $us$ (respectively $d s$) is $j$. \\
  \end{tabular}
\end{tabular}
\]
This notion will be used when $m\leq i$, so $l>0$ holds for each down
diagonal number $l$.

Note that if
$(us,d s)$ is correct up to $m$ w.r.t.\ $i$ and $cs$ (where  $m\leq i$)
then $(t l(us),[\myunderscore|d s])$ is correct up to $m$ w.r.t.\ $i+1$ and $cs$
(as
 the up diagonal number $l$ w.r.t.\ $i$ means the up diagonal number
$l-1$ w.r.t.\ $i+1$, for the down diagonal number $l$ this is $l+1$).

 Now the specification for ${\it p q s}$ is
 $S_{\it p q s} = S_{{\it p q s}1}\cup S_{{\it p q s}2}$ where
 \[
\begin{array}[b]{@{}l@{}}
S_{{\it p q s}1} =
    \left\{\,  p q s(i, cs, us, [\myunderscore|d s] ) \: \left|
          \begin{tabular}{l@{}}
                 $i>0$ \\
            $cs$ is correct up to $i$, \\
            $(us, d s)$ is correct up to $i$ w.r.t.\ $i$ and $cs$, \\
            terms $cs, us,d s$ are variable disjoint.
          \end{tabular}
          \right\}\right.
\vspace{1ex}
\\
    S_{{\it p q s}2} = 
      \big\{\, {\it p q s}(0, cs, us, d s) \mid cs, us, d s 
                                \mbox{ are distinct variables} \,\big\} %
\end{array}
\]
and the whole specification for \nqueens is%
\footnote{%
As a specification for the whole original program one can use
$S\cup S_{\it g l}\cup S_{\it g u}$, where
  \[
     \begin{array}[t]{@{}l@{}}
       S_{\it g l} =
       \big\{\, g l(i,[\seq[i]v]) \mid i\geq0, \ 
                                   \seq[i]v \mbox{ are distinct variables} 
                                   \,\big\},
      \vspace{1.5ex minus.8ex}
      \\      %
       S_{\it q u} = \left\{\, q u(i,c s) \:\left|\
        \begin{tabular}{@{}l@{}}%
            $i\geq0$, $c s$ is a list of length $i$,\\
             its members are $1,\ldots,i$,\\      
                  their up (down) diagonal numbers are distinct \\
        \end{tabular}
       \right.\right\}.
     \end{array}
  \]
} %
  \[
      S = S_{\it p q} \cup S_{\it p q s}.
  \]

For a specification to be useful, it should imply the program property of
interest.  (Each program is correct w.r.t.\ \TB, but this implies nothing.)
We now show that our specification captures the fact that the program solves
the $n$ queens problem.  Assume \nqueens is correct w.r.t.\ $S$
and consider the initial query
$Q ={\it p q s (n, q s_0, \myunderscore,\myunderscore)}$
from Section \ref{sec.program}, where $n>0$ 
and $q s_0$ is a list of variables of length $n$.
Any answer for $Q$ is a result $A\theta$ of unification of $Q$ and an atom
 $A\in \OO(\mbox{\sc \nqueens}) \subseteq S$.  
So the second argument of the answer, ${\it q s}_0\theta$, 
is a solution to the problem, as it is
a list of distinct members $1,\ldots,n$ with distinct up (and down)
diagonal numbers.

Here is a detailed justification.
As $A\in S_{\it p q s}$ and $A$, $Q$ are unifiable,  
$A= p q s(n, cs, us, [\myunderscore|d s] )$, 
where $cs$ is correct up to $n$.  
The length of open list $cs$ is $n$, because
the length is $\!{}\geq n$ (as $cs$ has members $1,\ldots,n$),
and is  $\!{}\leq n$ (as $cs$ is unifiable with a list of length $n$).
Thus the list $cs\theta = q s_0\theta$ is a permutation of $[1,\ldots,n]$.
The up (down) 
diagonal numbers of $1,\ldots,n$ in $q s_0\theta$ are those in $cs$, thus
distinct. 
{\sloppy\par}

Note that the specification is approximate (formally, that it is a proper
superset of the s-semantics of \nqueens).  For instance it allows multiple
occurrences of an element in  $us$ or $d s$ (in $S_{{\it p q s}1}$),
and does not require that $us$, $cs$ are open lists.
Note also that $S$ is closed under renaming.

\subsection{Correctness proof for \nqueens}
\label{sec.correctness.proof}
The proof of correctness of \nqueens w.r.t.\ $S$ is based on Theorem
\ref{theorem.correctness}.
The proof for the unary clauses
\[
\begin{array}{l}

    p q(I,[I|\myunderscore],[I|\myunderscore],[I|\myunderscore]). \\
    p q s(0,\myunderscore,\myunderscore,\myunderscore).  \\
\end{array}
\]
is immediate, as both are members of $S$ (and hence any their variants are).

Consider clause (\ref{clause4}):
\[
    p q(I,[\myunderscore|Cs],[\myunderscore|{\it Us}],[\myunderscore|D s]) \gets
            p q(I,Cs,{\it Us},D s). 
\]
It is easy to check that unifying the body of (\ref{clause4}) with any atom
from $S$ (thus from $S_{p q}$) and applying the mgu to the head of
(\ref{clause4}) results in an atom from $S_{p q}$,
provided that the clause and the atom are variable disjoint.
Hence $\Tpi[\!\{(\ref{clause4})\}](S)\subseteq S_{p q}\subseteq S$.

The nontrivial part of the proof is to show that
$\Tpi[\!\{(\ref{clause2})\}](S)\subseteq S$.
Remember that clause (\ref{clause2}) is
\[
\begin{array}{l}
    p q s(s(I),Cs,{\it Us},[\myunderscore|D s]) \gets
            p q s(I,Cs,[\myunderscore|{\it Us}],D s), \,
            p q(s(I),Cs,{\it Us},D s).
\end{array}
\]
Let $H$ stand for the head of the clause, and $B_1,B_2$ for its body atoms.
To find $ \Tpi[\!\{(\ref{clause2})\}](S)$ consider the unification of 
$B_1,B_2$ with a pair of atoms
\[
A_1={\it p q s}(i, cs_1, us_1, d s_1)\in S_{\it p q s}
\mbox{ \ and \ }
A_2={\it p q}(v, cs_2, us_2, d s_2) \in S_{\it p q}.
\]
(where $A_1$, $A_2$, $(H\mathop\gets B_1,B_2)$ are variable disjoint
and $i\geq0$).
Note that $us_1$ is of the form $[t|t']$, as
there are $i$ distinct up
diagonal numbers (in $cs_1$ w.r.t.\ $i$) and each is
$\geq 2{-}i$, hence some of them must be positive.

We have to show that
if $(B_1,B_2)$ and $(A_1,A_2)$ are unifiable then applying the mgu to $H$ 
results in a member of $S$.  So assume they are unifiable.
It is sufficient to consider a single mgu of $(B_1,B_2)$ and $(A_1,A_2)$
(as $S$ is closed under renaming).

We perform the unification in two steps, first unifying 
$(B_1,s(I))$ and $(A_1,v)$, then the remaining arguments of $B_2$ and $A_2$.
(Formally, Lemma 2.24 of \cite{Apt-Prolog} is applied here.)
For $i>0$ the first step produces
$\varphi=\{I/i,Cs/cs_1,\myunderscore/h,{\it Us}/{\rm t l}(us_1),
\linebreak[3]D s/d s_1, v/s(i) \}$
 (where $h$ is the head of $us_1$).
For $i=0$ we obtain
$\varphi=\{I/0,\linebreak[3]Cs/cs_1,\linebreak[3]{\it us}_1/[\myunderscore|{\it Us}],
\linebreak[3]D s/d s_1, v/s(0) \}$.
{\sloppy\par}

We show that pair
 $({\it Us},D s)\varphi$
 is correct up to $i$  
w.r.t.\ $i+1$ and $Cs\varphi= cs_1$.
This holds vacuously for $i=0$;\linebreak[3] for $i>0$ it follows from 
 $({\it Us},D s)\varphi=(t l(us_1),d s_1)$ and
$(us_1,t l(d s_1))$ being correct up to $i$ w.r.t.\ $i$. 
Hence for any substitution $\psi$
\sloppy
  \begin{equation}
    \label{property1}
\mbox{
     $({\it Us},D s)\varphi\psi$
     is correct up to $i$ w.r.t.\ $i+1$ and ${\it Cs}\varphi\psi$.
}
  \end{equation}
\fussy   %

In the second step, 
the remaining three arguments of $B_2\varphi$ are to be unified with
those of  $A_2\varphi$, this means obtaining an mgu $\psi$ for
$({\it Cs,Us,D s})\varphi$ and
\[
\begin{array}{l}
(cs_2, us_2, d s_2) =
(\,[\seq[k]c,s(i)|c_0],\, [\seq[k]u,s(i)|u_0],\, [\seq[k]d,s(i)|d_0]\,),
  \\
\mbox{%
where $k\geq0$, and
  $c_0,\ldots,c_k,\linebreak[3]u_0,\ldots,u_k,d_0,\ldots,d_k$ are distinct
  variables.}
\end{array}
\]
This gives $\varphi\psi$ as an mgu of $B_1,B_2$ with $A_1,A_2$.
As the terms 
$Cs\varphi,{\it Us}\varphi, D s\varphi,\linebreak[3] cs_2, us_2,
 d s_2
$
are variable disjoint, unifier $\psi$ can be represented as a union of three
 substitutions
\vspace{-.5ex}
\[
\begin{array}{c}
  \psi = \psi_{\rm c} \cup \psi_{\rm u} \cup \psi_{\rm d},
      \makebox[0pt][l]{\qquad where}
\qquad\qquad
 \\
Cs\varphi\psi = Cs\varphi\psi_{\rm c}, \quad
{\it Us}\varphi\psi = {\it Us}\varphi\psi_{\rm u}, \quad
D s\varphi\psi = D s\varphi\psi_{\rm d}, 
\end{array}
\]
and $\psi_{\rm c}, \psi_{\rm u}, \psi_{\rm d}$ are variable disjoint.
Hence $Cs\varphi\psi$, ${\it Us}\varphi\psi$ and $D s\varphi\psi$ are variable
disjoint.

Note that $Cs\varphi\psi$ is a g.v.d.\ (as the result of unification of two
variable disjoint
g.v.d.'s with disjoint sets of ground members), and its ground members are 
$1,\ldots,s(i)$.

In the rest of this proof we consider diagonal numbers in
${\it Cs}\varphi\psi$ w.r.t.\ $i+1$.

To show that $Cs\varphi\psi = cs_1\psi$ is correct up to $i+1$,
it remains to show that for $i>0$ the up (respectively down) diagonal numbers of
$s(0),\ldots,s(i)$ are distinct.
The up (resp.\ down) diagonal number for $s(i)$ is $k+1$, and the $k{+}1$-th
element of {\it Us} (resp.\ ${\it D s}$) is $s(i)$. 
So by (\ref{property1}) no up (down) diagonal number of $s(0),\ldots,i$ 
is $k+1$.
Moreover the up (down) diagonal numbers of $s(0),\ldots,i$ are distinct
(as $Cs\varphi$ is correct up to $i$).

To show that $({\it Us},{\it D s})\varphi\psi$ is correct up to $i+1$ 
w.r.t.\ $i+1$ and ${\it Cs}\varphi\psi$, it 
remains to show that for each each $j\in\{s(0),\ldots,s(i)\}$
the condition on the up (down) diagonal numbers
from the definition holds. 
For $j\leq i$ this follows from (\ref{property1}). For $j=s(i)$ this holds, as
$s(i)$ is the $k{+}1$-th member of 
${\it Cs}\varphi\psi$, ${\it Us}\varphi\psi$, and $D s\varphi\psi$,
and this $k+1$ is its up (and down) diagonal number.

Now applying the mgu to the head $H$  of the clause results in
$H\varphi\psi=
p q s(s(i), Cs\varphi\psi,{\it Us}\varphi\psi,[\myunderscore|D s\varphi\psi])$.
From what was shown above, by the definition of $S_{{\it p q s}1}$,
it follows that $H\varphi\psi\in S$.
We showed that $ \Tpi[\!\{(\ref{clause2})\}](S)\subseteq S$.
This completes the correctness proof.
{\sloppy\par}

\subsection{Comments}
\label{sec.comments.correctness}
%
% \enlargethispage{3.1ex}

The proof above can be compared with a correctness proof for \nqueens
\cite{drabent.nqueens.tplp.pre}
based on Herbrand interpretations and the standard semantics of definite
logic programs.  The specification used there is an Herbrand interpretation
(a set of ground atoms) and program correctness means that the least Herbrand
model of the program is a subset of the specification.

A difficulty had to be overcome, as some answers of \nqueens have instances
which are in a sense wrong.  For example, elements of $S_{{\it p q s}1}$
have ground instances in which the same queen is placed in two columns.
The main idea of solving the difficulty was to allow 
(i.e.\ to include in the specification) all ground atoms
$p q s(i, cs, us, [t|d s])$
in which $cs$ is not a list of distinct members. 
Thus the specification 
neglects the atoms with such $cs$ and describes the other arguments of 
$p q s$ only when $cs$ ``makes sense'', i.e.\ is a list with distinct members.
This outline is superficial, see \cite{drabent.nqueens.tplp.pre} for details.

The reader may compare the proof based on Herbrand interpretations with the
one presented here.  The former turns out substantially simpler.
Note that the presentation of the former proof in
{\cite{drabent.nqueens.tplp.pre}} is more detailed than that 
of Section \ref{sec.correctness.proof}
 where many details were skipped. 
For instance we have not proved that (under the given conditions)
unification of two g.v.d.'s results in a g.v.d.
Despite of this, the proof of Section \ref{sec.correctness.proof} above
is longer and seems more complicated.

The author began with a correctness proof based on the s-semantics, before it
turned out that employing the standard semantics was preferable.

\enlargethispage{2.1ex}

A well founded comparison of the volume of the two proofs could be obtained
by formalizing 
the specifications and the proofs, using some proof assistant.
This is however outside of the scope of this work.

One cannot claim that the proof presented here cannot be simplified.
The author may have missed some improvements.
It may be possible to find a more suitable specification which would
simplify the proof.  Possibly a toolbox of theorems dealing with properties of
substitutions, their composition, and the unification may help making our
proof smaller.  

Note however that the former proof employs simpler mathematical objects.
 Basically it
    deals only with ground atoms, sets of ground atoms,
    and inclusion of such sets.  
    Here we have to work with arbitrary atoms, variables, substitutions,
    and unification.
Hence 
it seems unlikely that a correctness proof based on
s-semantics can be made not more complicated than that employing the standard
semantics.

\section{Completeness}
\label{sec.completeness}
\subsection{Specification for completeness}
Obviously, program \nqueens is not complete w.r.t.\ specification $S$.
To construct a specification for completeness for \nqueens,
we need to describe (a set of)
atoms from $S_{p q s}$ which actually are answers of the program.

We first introduce some auxiliary notions.
Let us say that a g.v.d.\ $s=[\seq t|v]$ is {\bf short} if $t_n$ is a ground
term, or $n=0$.
  Consider the short g.v.d.\ $s$ and a $k\in\{1,\ldots,n\}$ such that
$t_k$ is ground and   $t_{k+1},\ldots,t_{n-1}$ are variables;
if all $\seq[n-1]t$ are variables then let $k=0$.
Now the g.v.d.\ $s$ {\bf with} $t_n$ {\bf removed} is $s'=[\seq[k]t|v]$.
For an $i\in\{1,\ldots,n-1\}$, the g.v.d.\ $s$ {\bf with} a ground $t_i$ 
{\bf removed} is
obtained from $s$ by replacing  $t_i$ by a new variable.
Note that in both cases a short g.v.d.\ with a ground member removed is a
short g.v.d.

Now this is our specification for $p q s$ for completeness:
\[
S_{\it p q s}^0 = 
\left\{\,  p q s(i, cs, us, [\myunderscore|d s] )  \left|
      \begin{tabular}{@{\,\,}l}
             $i>0$, \ %
        $cs$ is correct up to $i$, \\
        $(us, d s)$ is correct up to $i$ w.r.t.\ $i$ and $cs$, \\
        terms $cs, us,d s$ are variable disjoint, \\
        $cs, us,d s$ are short g.v.d.'s, \\
                    if $j$ is a ground member of $us$ or $d s$ \\
                    then
                    $j\in\{1,\ldots,i\}$,   \\
                    if $j$ is a ground member of $us$ then its\\
                    up diagonal number in $cs$ w.r.t.\ $i$ is ${>}\,0$ \\
      \end{tabular}
      \right\}\right..
\]
So here we require $us$ and $d s$ to have only such ground members that are
necessary for correctness of $(us,d s)$.
Note that $S_{\it p q s}^0\subseteq S_{\it p q s}$.

Note that such specification makes sense, as its atoms describe all the
solutions to the $i$ queens problems.
So completeness of \nqueens (w.r.t.\ $S_{\it p q s}^0$)
implies that each solution is contained 
in an answer to the initial query considered previously.

We are interested in completeness of \nqueens w.r.t.\ $S_{\it p q s}$.
However this cannot be proved using Theorem \ref{theorem.completeness}.
We need to strengthen the specification, to describe requirements on $p q$
and on the answers for $p q s$ with the first argument $0$.  
Fortunately, relevant fragments of the specification for correctness can be
reused here.
Now our specification for completeness of \nqueens is
\[
S^0 = 
S_{\it p q s}^0\cup S_{{\it p q s}2}\cup S_{\it p q}.
\]
Note that this is a proper subset of the specification for correctness $S$.

\subsection{Completeness proof}

Now we apply Theorem \ref{theorem.completeness} to
 prove completeness of the program, i.e.\ that $S^0\subseteq\OO(\nqueens)$.
First 
let us define, 
similarly to {\cite{drabent.nqueens.tplp.pre}},
a level mapping $|\ |\colon S\to\NN$ by
\[
\begin{array}{l}
  | \, {\it p q s}(i,cs,us,d s) \,| = |i|+|cs|,
\\
  | \, {\it p q}(i,cs,us,d s) \,| = |cs|,
\end{array}
 \qquad \mbox{where} \qquad
    \begin{array}{l}
      |\, [h|t]\, | = 1+|t|,    \\
      |\, s(t)\, | = 1+|t|,    \\
      |f(\seq t)| = 0,         \\
      |v| = 0,
    \end{array}
\]
where $i,cs,us, d s, h,t,\seq t\in\TU$, $v\in\Var$  and
$f$ is any $n$-ary function symbol ($n\geq0$)
  distinct from $s$ and from $[\ | \ ]$. 
  Note that for an (open) list $l$, its length is $|l|$.
  Note also that
 if $s'$ is a short g.v.d.\ $s$ with a ground member removed then
 $|s'|\leq|s|$.

 The atoms from $S_{{\it p q s}2}$ and those of the form
 ${\it p q}(v,[v|\myunderscore],[v|\myunderscore],[v|\myunderscore])
\in S_{\it p q}$ are variants of unary clauses of \nqueens, thus obviously
the are in, respectively, 
$\Tpi[\{(\ref{clause1})\}](\emptyset)$ and
$\Tpi[\{(\ref{clause3})\}](\emptyset)$.

The nontrivial part of the proof is to show that the sufficient condition
from Theorem \ref{theorem.completeness} holds for the elements of
$S_{\it p q s}^0$.

Consider an atom $A = {\it p q s}(s(i), cs, us, [v|d s]) \in S_{\it p q s}^0$. 
Let $j$ be the (both up and down) diagonal number of $s(i)$ 
in $cs$ w.r.t.\ $s(i)$.  So $s(i)$ is the $j$-th member of each
short g.v.d.'s $cs,us,d s$.
We show that $A\in \Tpi[\{(\ref{clause2})\}](\{A_1,A_2\})$,
for certain $A_1,A_2\in S^0$.  
Remember that clause (\ref{clause2}) is
\[
\begin{array}{l}
    p q s(s(I),Cs,{\it Us},[\myunderscore|D s]) \gets
            p q s(I,Cs,[\myunderscore|{\it Us}],D s), \,
            p q(s(I),Cs,{\it Us},D s).
\end{array}
\]
Below we can assume that  $(\ref{clause2}),A_1,A_2$ are
variable disjoint,
if necessary $A_1$ or $A_2$ can be replaced by its suitable variant.

As $A_2$ we
 choose $A_2 ={\it p q}(v', cs'', us'', d s'')\in S_{\it p q}$, 
where $v'\in\Var$ is the \mbox{$j$-th} member of each  $cs'', us'', d s''$.
For $i=0$ we choose $A_1 = {\it p q s}(0, v_1,v_2,v_3)\in S_{{\it p q s}2}$. 
Let $\rho=\{v'/s(0)\}$.
A most general unifier of $A_1,A_2$ and the body of clause (\ref{clause2}) 
is  $\theta=\rho\cup
     \{ I/0,{\it Cs/cs''\!\rho, Us/us''\!\rho, D s/d s''\!\rho},\ldots \}$.
Note that $cs''\!\rho, us''\!\rho, d s''\!\rho$ are short g.v.d.'s.
Applying $\theta$ to the head of the clause results in    
$ {\it p q s}(s(0),{\it cs'', us'', [\myunderscore|d s'']})$.
This is a variant of $A$. 
Thus  $A\in \Tpi[\{(\ref{clause2})\}](\{A_1,A_2\})$,
{\sloppy\par}

If $i>0$ then
as $A_1$ we choose $A_1 = {\it p q s}(i, cs', [t|us'], d s')$, 
where
$cs'$ (respectively $us',\, d s'$) is $cs$ ($us$,\ $d s$) with $s(i)$ removed, 
and $t$ is as follows.  If 1 is the up diagonal number in $cs$
w.r.t.\ $i$ of some $k\in\{s(0),\ldots,i\}$ then $t=k$.  Otherwise $t$ is a
variable such that  $A_1$ is linear.

Note that the diagonal numbers of $1,\ldots,i$ in $cs$ are the same as those
in $cs'$.  
So a pair is correct up to $i$ w.r.t.\ $k$ and $cs$
iff it is correct up to $i$ w.r.t.\ $k$ and $cs'$ (for any $k\geq i$).

We first show that $A_1\in S_{\it p q s}^0$.
Note that $cs', us',d s'$ are short g.v.d.'s.
They are variable disjoint, as $cs,us,d s$ are.
Also, $cs'$ is correct up to $i$ (as $cs$ is correct up to $s(i)$), and
$(us',d s')$ is correct up to $i$ w.r.t.\ $s(i)$ and $cs$ (as $(us,d s)$ is,
up to $s(i)$).
Thus $([t|us'], {\rm t l}(d s'))$ is correct w.r.t.\ $i$ and $cs'$ up to $i$.
A ground member $m$ of $[t|us']$ or of ${\rm t l}(d s')$
is $k$ or a member of $us'$ or $d s'$. Hence $m\in\{1,\ldots,i\}$.

It remains to show that the diagonal numbers of the ground members of
$[t|us']$ w.r.t.\ $i$ are positive.
In this paragraph we consider diagonal numbers and correctness w.r.t.\ $cs'$,
so we skip the phrase ``w.r.t.\ $cs'$''.
Consider a ground member $m$ of $[t|us']$.  If $m=t$ then its 
up diagonal number w.r.t.\ $i$ is 1.  If $m$ is a member of $us'$ then
$m\neq s(i)$ and $m$ is a member of $us$.
As $us$ is the third argument of ${\it p q s}$ in $A\in S_{\it p q s}^0$, 
$m\in\{1,\ldots,i\}$ and 
the up diagonal number of $m$ w.r.t.\ $i+1$ is positive.
Thus the up diagonal number of $m$ w.r.t.\ $i$ is ${}>1$.
This completes a proof that  $A_1\in S_{\it p q s}^0$.

Now we show that $A_1,A_2$ are unifiable with the body atoms $B_1,B_2$ of the
clause (\ref{clause2}) and the resulting mgu produces (a variant of) $A$.
Similarly as in the previous proof, let us perform unification in two steps.
Unifying $(A_1,v')$ with $(B_1,s(I))$  results in
$\varphi=\{I/i,Cs/cs',\myunderscore/t,{\it Us}/us',\linebreak[3]
D s/d s', v'/s(i)\}$.
The rest of unification is unifying three variable disjoint short g.v.d.'s
$cs',us',d s'$ with three short g.v.d.'s $cs''\varphi,us''\varphi,d s''\varphi$,
the latter are $cs''\{v'/s(i)\},\,us''\{v'/s(i)\},\linebreak[3]d s''\{v'/s(i)\}$
(as $v'$ is the only variable from $\varphi$ that occurs in $cs'',us'',d s''$).
Remember that $cs'$ is $cs$ with its $j$-th member $s(i)$ removed, and
$cs''\{v'/s(i)\}$ is a short g.v.d.\ with its $j$-th member $s(i)$, and this
is the only nonground member of the g.v.d.  Hence unifying $cs'$ and 
$cs''\{v'/s(i)\}$ results in $cs$.  The same holds for $us'$ and $d s'$.
Applying the resulting mgu of $A_1,A_2$ and $B_1,B_2$ to the head of the
clause results in A.

This completes our proof that 
$A\in \Tpi[\{(\ref{clause2})\}](\{A_1,A_2\})$, where $A_1,A_2\in S^0$.
Note now that $|A_2|=|cs''|=j$.  For $i=0$, $|A_1|=0$; for $i>0$ we have
$|A_1| = i + |cs'| \leq i + |cs|$
(as $cs'$ is the short g.v.d.\ $cs$ with a ground member removed).
Also, $A= i+1+|cs| \geq i+1+j$ (as the g.v.d.\ $cs$ has at least $j$ members).
Hence
 $|A|>|A_1|$ and $|A|>|A_2|$. So we have shown that
the sufficient condition for completeness from Theorem \ref{theorem.completeness}
holds for any $A\in S_{\it p q s}^0$.

It remains to show that the sufficient condition holds for any atom
$ B_k = p q(\,v,\, [\seq[k]c,v|c_0],\, [\seq[k]u,v|u_0],\, [\seq[k]d,v|d_0] \,) 
\in S_{p q}$,
where $k>0$. Note that $|B_k|=k+1$. 
We skip (simple) details of showing that 
$B_k'\in \Tpi[\{(\ref{clause4})\}](B_{k-1})$ for some variant $B_k'$ of $B_k$.
This completes the proof.
{\sloppy\par}

\subsection{Comments.}
Similarly as in Section \ref{sec.comments.correctness}, the completeness
proof above can be compared with one using the standard semantics
\cite{drabent.nqueens.tplp.pre}.  First note that here we had to use a
substantially more complicated specification.  In the former work, the
specification for completeness could contain only some ``meaningful'' ground
instances of the answers (and lot of other ones have been skipped).  So that
specification is a rather small subset of the least Herbrand model of
\nqueens.  Here, in the context of the s-semantics, each element of the
specification has to be an exact answer for a most general query.
Describing this is rather tedious.%
\footnote{%
It seems that our specification $S^0$ equals to $\OO(\nqueens)$.
Checking this hypothesis is irrelevant for the main purpose of this paper.
Correctness w.r.t.\ $S$ is sufficient for the program solving the $n$ queens
problem.  So we do not need to discuss correctness w.r.t.\ any stronger
specifications, for instance $S^0$.
}

Also the completeness proof itself is larger than that based on the standard
semantics.  Additionally, as in the case of correctness, the proof here is
presented in a less detailed way than that in \cite{drabent.nqueens.tplp.pre}.
Other comments on comparing the correctness proofs from 
Section \ref{sec.comments.correctness} apply also here.

\section{Summary}

This paper presents correctness and completeness proofs (together with 
suitable specifications) of program \nqueens.
It is a definite clause program, working on non-ground terms.
The specifications and
proofs are based on the s-semantics
\cite{DBLP:journals/tcs/FalaschiLPM89,s-semantics94,DBLP:journals/tcs/Bossi09}.
The employed approach is declarative;  the specifications\,/\,proofs
abstract from any operational semantics.  Our specification is approximate,
it consists of separate specifications for correctness and completeness.

The proposed sufficient conditions for correctness and completeness seem to
be a contribution of this work. 
The employed simplification of the s-semantics may be of separate interest.
The author is not aware of any published examples of applying the s-semantics
to reasoning about properties of particular programs.

The program works on nonground data, and the s-semantics explicitly deals
with variables in program answers.  Thus the choice of this semantics
seems reasonable.  
However comparison with analogical specifications and proofs
 \cite{drabent.nqueens.tplp.pre}  based on the
standard semantics and (ground) Herbrand interpretations
shows that the latter are simpler.  
This, perhaps surprisingly, disproves a hypothesis about suitability of
s-semantics for such cases.

\bibliographystyle{alpha}
\bibliography{bibshorter,bibpearl,bibmagic,bibs-s}

\end{document}